\documentclass[aip,graphicx]{revtex4-1}
\usepackage[cp1251]{inputenc}
\usepackage[T2A]{fontenc}
\usepackage[english]{babel}
\usepackage{amssymb,latexsym,amsmath,amscd}
\usepackage{graphicx,color,framed}
\begin{document}
\selectlanguage{english}
\newcommand{\angstrom}{\text{\normalfont\AA}}
\title{Generation of mechanical force by grafted polyelectrolytes in an electric field. Application to polyelectrolyte-based nano-devices}
\author{\firstname{N.~V.}~\surname{Brilliantov}}
\email{nb144@le.ac.uk}
\affiliation{Department of Mathematics, University of Leicester, Leicester LE1 7RH, United Kingdom}
\author{\firstname{Yu.~A.}~\surname{Budkov}}
\affiliation{Department of Applied Mathematics, National Research University Higher School of Economics, Moscow, Russia}
\author{\firstname{C.}~\surname{Seidel}}
\affiliation{Max Planck Institute of Colloids and Interfaces, Science Park Golm, D-14424 Potsdam, Germany}
\begin{abstract}
We analyze theoretically and by means of molecular dynamics (MD) simulations the  generation of mechanical force by a polyelectrolyte (PE) chain grafted to a plane. The PE is exposed to an external electric field that favors its adsorption on the plane.  The free end of the chain is linked to a deformable target body. Varying the field one can alter the length of the non-adsorbed part of the chain. This entails variation of the deformation of the target body and hence variation of the arising in the body force. Our theoretical predictions for the generated force are in a very good agreement with the MD data. Using the developed theory for the generated force we study the effectiveness of possible PE-based nano-vices, comprised of two clenching planes connected by PEs and exposed to an external electric field. We exploit Cundall-Struck solid friction model to describe the friction between a particle and the clenching planes. We compute the self-diffusion coefficient of a clenched particle  and show that it drastically decreases even in weak applied fields. This demonstrates the efficacy of the PE-based nano-vices, which may be a a possible alternative to the existing nano-tube nano-tweezers and optical tweezers.
\end{abstract}
\maketitle
\section{Introduction}
Future nanotechnology will use molecular devices executing various manipulations with nano-size objects, such as colloidal particles, vesicles, macromolecules, viruses, small bacteria, cell organelles, etc. Such objects, however, can not stay at rest on their own due to heat, manifesting itself in a form of thermal fluctuations. To execute highly precise manipulations with these objects one needs to keep them immobilized. Therefore,  the devices that can clench and unclench nano-size objects under an external control will be massively demanded. These devices, which can be termed as "nano-vices" or "nano-nippers" should be able to operate in solutions, including aqueous solutions, where manipulations with biological nano-size objects are expected. Recently, nano-tube nano-tweezers operated by an electric field, that can clench a nano-size object, has been proposed~\cite{Nanotweez99}. Such devices, however, can not work effectively in aqueous solutions; moreover, the range of forces, as well as the range of operating distances, is relatively narrow for such devices~\cite{Nanotweez99}.  Optical tweezers require specific optical properties of an immobilized object which also restricts their applications~\cite{MacDonald2002}. Therefore it seems reasonable to consider  polyelectrolyte-based nano-vices, operated by an electric field, which could be a very promising alternative to existing nano-tube nano-tweezers and optical tweezers. These can operate in aqueous solutions and demonstrate a wide range of operating forces and distances. They also overcome limitations imposed by the optical properties of nano-objects~\cite{MacDonald2002}.

In very simple terms the  nano-vices may be comprised of a few charged polymer chains the so-called polyelectrolytes (PE) that are linked to two surfaces and exposed to an electric field that serves as a control signal. Varying the field one can alter the length of the non-adsorbed part of the chains and hence the distance between the planes, see Fig.1. In this way one can clench and unclench a nano-size object (target body) placed between the planes.

\begin{figure}
\center{\includegraphics[width=0.7\linewidth]{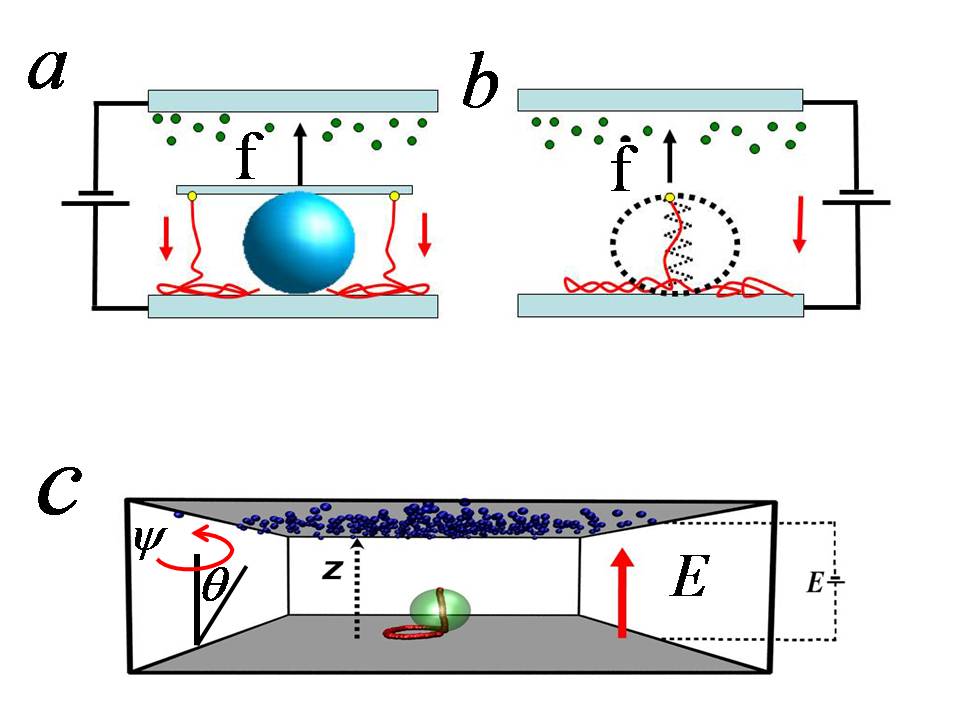}}
\caption{(a) Charged polyelectrolyte (PE) chains are linked to two planes. They are attracted to the one, oppositely charged plane, which causes the chains' contraction, as indicated by the downwards arrows. This leads to the clenching of the target body (e.g. a colloidal particle, virus, etc.) and  gives rise to the restoring force $f$, indicated by the upward arrows. (b) In the numerical simulations the target body is modelled by a spring with a given force-deformation relation, which is linked to the free end of the PE chain. (c) Typical simulation snapshot of a PE chain in the electric field perpendicular to the charged plane (shown by the upward arrow) and linked to the target body. Illustration of the Cundall-Struck model.}
\label{fig1}
\end{figure}

\begin{figure}
\center{\includegraphics[width=0.7\linewidth]{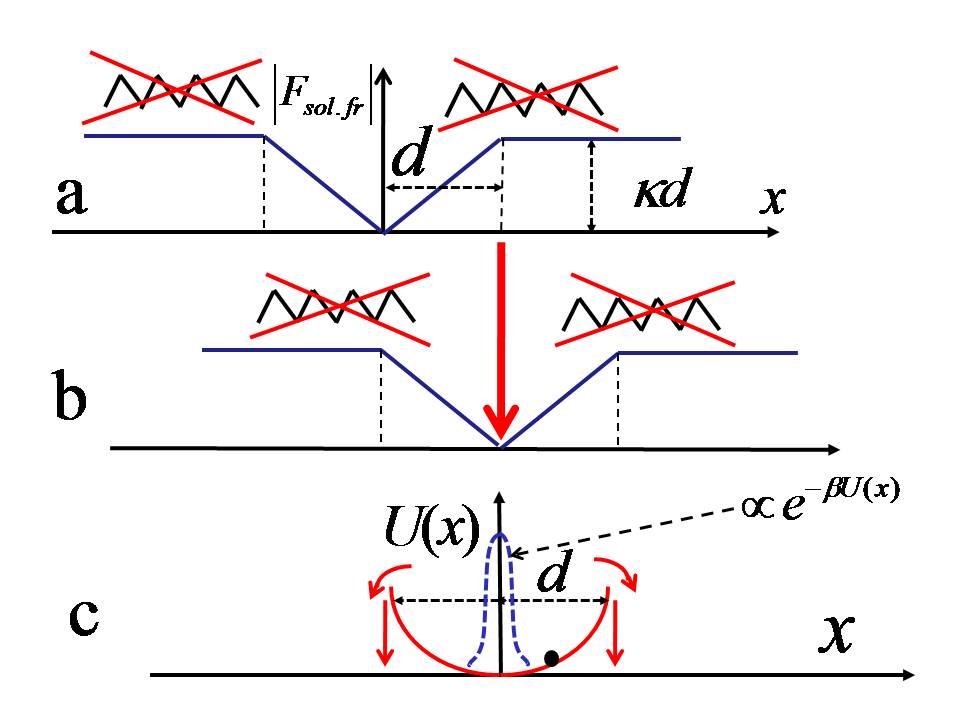}}
\caption{(a) Solid friction force grows linearly with the displacement $x$ as for a linear spring. (b) When $x$ exceeds $d$, the "spring" breaks and particle moves to a new potential well, shifted by $d$. (c) Particle's positions in the potential well, associated with the solid friction force, obey the equilibrium distribution.}
\label{fig2}
\end{figure}

Recent technological achievements, e.g. the production of nano-sheets~\cite{Nanosheets2011} enables a practical realization of such devices, hence it is very important to develop a theory of the respective devices and quantitatively describe the basic physical processes there. Hence one needs to (i) develop a theory of a conformational response of a PE, with one end linked (grafted) to a charged surface and the other one to a deformable target body, to a varying electric field  and (ii) quantify the ability of nano-vices to clench a nano-size object.

The response of PEs to external electric fields has been extensively studied the last few decades~\cite{Boru98,Joanny98,Dobry2000,Borisov2001,Netz2003,Muthu2004,Bril_nanov1,Bril_nanov2,Bril_nanov3}, including the adsorption of PEs on oppositely charged surfaces of different geometries~\cite{MetzlerCherst:2014,MetzlerCherst2015}.
In particular, the conformational response of a PE in an electric field, under an action of
a constant force, has been analysed in Ref.~\cite{Joanny98}.  In the context of nano-vices, however,  this problem has been addressed only recently~\cite{Bril_nanov1,Bril_nanov2,Bril_nanov3}. Moreover, the quantification of the clenching ability of PE-based nano-vices has not been studied yet. In Ref.~\cite{Bril_nanov1} a theory of the phenomenon, based on a model "physical" approach has been developed for the case of a constant force acting on a free (non-grafted) end of a PE. Later, in Ref.~\cite{Bril_nanov2}, a generalization of this theory for the case of force depending on deformation has been reported. In Ref.~\cite{Bril_nanov3} a first-principle theory of this phenomenon has been elaborated; all the theoretical studies were accompanied by extensive molecular dynamics (MD) simulations~\cite{Bril_nanov1,Bril_nanov2,Bril_nanov3}.

In the present paper we first discuss the theoretical approach of~\cite{Bril_nanov1,Bril_nanov2,Bril_nanov3} and  give a shorter and more straightforward derivation of the main result of the first-principle theory ~\cite{Bril_nanov3}. Then we analyse the efficacy of nano-vices calculating the self-diffusion coefficient of a clenched particle. In particular, we show that for rather weak  electric fields the PE-based nano-vices can effectively immobilize a clenched Brownian particle, reducing its self-diffusion coefficient by a few orders of magnitude.

\section{Conformation of a polyelectrolyte linked to a target body in an external electric field}
\subsection{Model} We consider  a system comprised of a chain of $N_0+1$ monomers, which is anchored to a planar surface at $z=0$. The anchoring end-monomer is uncharged. Each of the remaining $N_0$ beads carries the charge $-qe$ ($e>0$ is the elementary charge) and $N_0$ counterions of charge $+qe$ make the system neutral. For simplicity
we consider a salt-free solution. However it may be shown \cite{Budkov_salt} that for intermediate and strong electric fields, the presence of salt ions (up to physiological concentrations) leads to a simple renormalization of the external field, so that the qualitative nature of the phenomenon remains unchanged~\footnote{The salt coins  simply screen the adsorbing plane for such fields, but do not practically screen the bulk part of the chain.}. Hence a salt-free case, which allows an analytical treatment, is generic.

We use the freely joint chain model with the length of the inter-monomer link equal to $b$. The MD simulations~\cite{Bril_nanov1,Bril_nanov2,Bril_nanov3} justify the application of this model to the analysed phenomena.  We consider a salt-free system, so that no other microions present in the solution, which has the  dielectric permittivity $\varepsilon$. The chain is grafted (linked) to a charged plane so that the external electric field ${\bf E}$ acts perpendicular to the plane, favoring the chain adsorption. The free end of the PE is linked to a deformable body, modeled by a spring, Fig.1(b).

The deformation energy of the spring $U_{\rm sp}=U_{\rm sp}(h- h_0)$ depends on the deformation $h-h_0$, where $h_0$ and $h$ are the sizes of an undeformed and deformed target body respectively. The restoring force acting on the free end of the chain then reads,
\begin{equation}
\label{eq:Uspgen}
U_{\rm sp}=U_{\rm sp}(h- h_0) \quad  \qquad f=-\frac{\partial}{\partial h} U_{\rm sp}(h- h_0).
\end{equation}
In Refs.~\cite{Bril_nanov1,Bril_nanov2,Bril_nanov3} the following dependencies have been studied in detail theoretically and numerically: $U_{\rm sp}=fh$, $U_{\rm sp}=(1/2)\kappa (h-h_0)^2$, $U_{\rm sp}=(\kappa/\gamma) |h-h_0|^{\gamma-1}$ and $U_{\rm sp}=(2/5)\kappa (h-h_0)^{3/2} \theta(h_0-h)$, which refer accordingly to the constant force, linear, non-linear and Hertzian spring. Here $\kappa$  is the elastic constant of the spring and $\theta(x)$ is the Heaviside step function.

For the values of the external field and system parameters addressed in the present study, the counterions are practically decoupled from the chain, being accumulated near the upper plane, see Fig.1(c). Hence the interaction of the chain with the external field as well as interactions between the chain monomers are not screened by the counterions. To find the conformational response of the PE to the external electric field and the force arising in the target body, one needs to find the free energy of the system and minimize it with respect to relevant parameters.

\subsection{Free energy of the chain}
We first calculate a conditional free energy of the chain $F_{\rm ch}$ with the following conditions imposed: (i) $N$ monomers of the chain are desorbed and $N_s=N_0-N$ are adsorbed on the charged plane; (ii) the distance between the free chain end, linked to the target body, and the grafting plane is $z_{\rm top}$ and (iii) the end-to-end distance of the adsorbed part of the chain is ${\bf R}$. Minimizing the conditional free energy $F_{\rm ch}(N, z_{\rm top}, {\bf R})$ with respect to these variables one can  find the equilibrium values of $N$, $z_{\rm top}$ and ${\bf R}$; using these quantities we then compute the force acting on the target body.

For the freely joint model the location of all monomers of the chain are determined by $N_0$ vectors ${\bf b}_i= {\bf r}_i- {\bf r}_{i+1}$, which join centers of $(i+1)$-st  and $i$-th monomer; each of these vectors has the same length $b$. We enumerate the monomers starting from the free end linked to the target body. Then the beads with numbers $1,2,\ldots N$ refer to the bulk part of the chain, while with numbers $N+1, N+2, \ldots N_0$ to the surface part. The $N_0+1$st neutral bead, located at the origin, ${\bf r}_{N_0+1}=0$, is anchored to the surface. We assume that the adsorbed part of the chain forms a flat structure, so that the centers of the adsorbed beads lie at the plane $z=0$. In other words, we ignore the off-surface loops of this part of the chain. Then the location of the $k$th bead of the chain and the  distance between centers of $i$th and $j$th bead read,
\begin{equation}
\label{eq:rij}
{\bf r}_k = \sum_{s=k}^{N_0} {\bf b}_{s} \qquad \qquad
{\bf r}_{ij} = \sum_{s=i}^{j} {\bf b}_s
\end{equation}
The orientation of each vector ${\bf b}_s$ is characterized by the polar $\theta_s$ and azimuthal $\psi_s$ angles, where the axis $OZ$ is directed perpendicular to the grafting plane, see Fig.1(c). Therefore the distance between the grafting plane at $z=0$ and the $k$th bead of the bulk part of the chain as well as the hight of the top bead for $k=1$ read:
\begin{equation}
\label{eq:rij}
{\bf r}_k = \sum_{s=k}^{N_0} {\bf b}_{s} \qquad \qquad
{\bf r}_{ij} = \sum_{s=i}^{j} {\bf b}_s
\end{equation}
\begin{equation}
\label{eq:ztop}
z_k= b \sum_{s=k}^N \cos \theta_s; \qquad \qquad z_{\rm top} = z_1=b \sum_{s=1}^N \cos \theta_s.
\end{equation}
The location of the top bead, linked to the target body determines the deformation energy of the target body and the according force which acts on the chain:
\begin{equation}
\label{eq:Uspztop}
U_{\rm sp} (z_{\rm top}) = U_{\rm sp} (z_{\rm top}-z_{\rm top, \, 0}), \qquad \quad f=-\frac{\partial U_{\rm sp}} {\partial z_{\rm top}} ,
\end{equation}
where $z_{\rm top, \, 0}$ is the coordinate of the top bead for the case when the target body is not deformed. Since the chain is not screened by the counterions, the potential associated with the external field $E$ is  $\varphi_{\rm ext} (z) = -Ez$. Hence the interaction energy of the bulk part of the chain with the external potential reads,

\begin{equation}
\label{eq:Ext}
H_{\rm ext} = \sum_{k=1}^N -qe\varphi_{\rm ext}(z_k)= bqeE \sum_{k=1}^N\sum_{s=k}^N \cos \theta_s = bqeE\sum_{s=1}^N s\cos \theta_s.  \nonumber
\end{equation}
The interaction energy with the external field  of the adsorbed part of the chain, located at $z=0$, is equal to a constant which we take equal to zero.

The electrostatic interactions between the chain monomers are also unscreened and may be written using the Fourier transform of a Coulomb potential $q^2 e^2 / \varepsilon r$ as~\cite{Bril_nanov3}
\begin{equation}
\label{eq:Hself}
H_{\rm self} = \frac12 \sum_{l=1}^{N_0}\sum_{m=1\,m\neq l}^{N_0} \frac{q^2 e^2}{\varepsilon r_{lm}} =\frac{q^2 e^2 }{2 \varepsilon}
\sum_{l\neq m } \int \frac{d {\bf k }}{(2\pi )^3} \left( \frac{4 \pi}{k^2} \right) e^{  i  \sum_{s=l}^{m} {\bf k} \cdot {\bf b}_s},
\end{equation}
where we exploit  Eq.~(\ref{eq:rij}). Taking into account that the end-to-end vector of the adsorbed part of the chain has the form ${\bf R} =\sum_{i=N+1}^{N_0}{\bf b}_i$, we can finally write the conditional free energy as
\begin{equation}
\label{eq:Fcond}
\beta F_{\rm ch}  (N, z_{\rm top}, {\bf R}) = - \log {\cal Z}_{\rm ch} (N, z_{\rm top}, {\bf R})
\end{equation}
where $\beta =(k_BT)^{-1}$ with $k_B$ and $T$ being respectively the Boltzmann constant and temperature and ${\cal Z}_{\rm ch}$ is the conditional partition function:
\begin{equation}
\label{eq:Zch}
 {\cal Z}_{\rm ch}=\int_{0}^{1} d\cos{\theta_{1}} \ldots \int_{0}^{1} d\cos{\theta_{N}} \int_{0}^{2\pi}d\psi_{1} \ldots \int_{0}^{2\pi}d\psi_{N_0}
  e^{-\beta U_{\rm sp} -\beta H_{\rm self} - \beta H_{\rm ext}}\nonumber
 \end{equation}
 \begin{equation}
\times b^3 \delta\left(\!\!z_{\rm top} -  b \sum_{s=1}^{N}\cos{\theta_{s}}\right)
\delta\left({\bf R} - \sum_{i=N+1}^{N_0}{\bf b}_i \right), \nonumber
\end{equation}
where the factor $b^3$ keeps ${\cal Z}_{\rm ch}$ dimensionless and $\psi_i$ and $\theta_i$ are respectively the azimuthal and polar angles of an inter-monomer vector ${\bf b}_i$ (see Fig.1(c)). We also take into account that $\theta_{N+1}= \theta_{N+2}\ldots =\theta_{N_0}=\pi/2$, since the adsorbed part of the chain forms a flat structure. Note that ${\bf R}$ is a two-dimensional vector on the plane and the vectors ${\bf b}_i$, for $i=N+1, \ldots N_0$ have zero $z$-component.

Using the integral representation for the delta function,
$$
\delta ({\bf r} ) = (2 \pi)^{-3}  \int  e^{i {\bf p} \cdot {\bf r}} d{\bf p}
=(2\pi)^{-1} \int_{-\infty}^{\infty} e^{i p_z z} dp_z \cdot
(2\pi)^{-2}  \int d{\bf p}_{\perp} e^{-i{\bf p}_{\perp} \cdot {\bf r}_{\perp}}
$$
where the vector ${\bf r}$ has a lateral and $z$ components,  ${\bf r}=({\bf r}_{\perp}, z)$, we recast Eq.~(\ref{eq:Zch}) into the form
\begin{equation}
\label{eq:Zch1}
{\cal Z}_{\rm ch} \frac{b^3}{(2\pi)^3} \int_{-\infty}^{\infty} dp_z e^{-ip_z z_{\rm top} -\beta U_{\rm sp}(z_{\rm top})}
\int_0^1 d\eta_1 \ldots \int_0^1 d\eta_N \, e^{\sum_{s=1}^N (ip_z-\tilde{E}) \eta_s}\nonumber
\end{equation}
\begin{equation}
\times\int d{\bf p}_{\perp} e^{-i{\bf p}_{\perp} \cdot {\bf R}} {\cal Z}_{\psi}
 \left<e^{-\beta H_{\rm self}} \right>_{\psi}, \nonumber
\end{equation}
where $\eta_s  =\cos \theta_s$ and we define
\begin{equation}
\label{eq:Zpsi}
{\cal Z}_{\psi} =\int_0^{2\pi} d \psi_1 \ldots \int_0^{2 \pi} d \psi_{N_0}
e^{i  {\bf p}_{\perp} \sum_{s=N+1}^{N_0} {\bf b}_s }
= (2 \pi)^{N} \prod_{s=N+1}^{N_0} \int_0^{2 \pi} e^{ip_{\perp}b \cos \psi_s} d \psi_s\nonumber
\end{equation}
\begin{equation}
=(2 \pi)^{N_0} \left[ J_0(p_{\perp}b)\right]^{N_s}. \nonumber
\end{equation}
Here we take into account that ${\bf p}_{\perp} \cdot {\bf b}_s =p_{\perp}b \cos \psi_s$, that $N_s=N_0-N$, and use the integral representation of the zeroth-order Bessel function, $J_0(x)= (2 \pi)^{-1} \int_0^{2\pi} \cos (x \cos \psi) d \psi$. We also define the averaging over the azimuthal angles $\psi_s$ as
\begin{equation}
\label{eq:avpsi}
\left< (\ldots) \right> =\frac{1}{{\cal Z}_{\psi}} \int_0^{2\pi} d \psi_1 \ldots \int_0^{2 \pi} d \psi_{N_0}  e^{i  {\bf p}_{\perp} \cdot  \sum_{s=N+1}^{N_0} {\bf b}_s } (\ldots) .
\end{equation}
Taking into account the long-range nature of the Coulomb interactions in $H_{\rm self}$, one can expect that the mean-field approximation will have a good accuracy. The mean-field approximation deals with the average quantities and neglects fluctuations. The average quantities are described by the first-order cumulants, while fluctuations by the higher order cumulants. Hence we adopt the following mean-field approximation:
\begin{equation}
\label{eq:MF}
\left< e^{-\beta H_{\rm self}} \right>_{\psi} \approx e^{-\beta \left< H_{\rm self}\right>_{\psi}}.
\end{equation}
Performing integrations over $\psi_1, \ldots \psi_{N_0}$ and then over ${\bf k}$, we arrive at (see e.g. \cite{Bril_nanov3} for detail, where similar quantities have been computed):

\begin{equation}
\label{eq:Hselfav}
\left< \beta H_{\rm self} \right>_{\psi} \!\!\!\!\simeq \!\! \!\! \frac{\tilde{l}_B}{2} \sum_{s_1\neq s_2}^N \frac{1}{B(s_1,s_2)}  +  \frac{\tilde{l}_B}{2} \sum_{s_1\neq s_2}^{N_s} \frac{ \, e^{p^2b^2|s_2-s_1|/8}} {\sqrt{ |s_2-s_1|}}  + \sum_{l=1}^{N}\sum_{m=1}^{N_s} \frac{\tilde{l}_B}{\sqrt{ B^{2}(l,N ) -\frac14 b^2p^2 m^2}},\nonumber
\end{equation}
where $l_B=e^2/(\varepsilon k_BT)$ is the Bjerrum length,  $\tilde{l}_B=l_Bq^2/b$ and $B(s_1,s_2) =\sum_{s=s_1}^{s_2} \eta_s$. Next, integration in Eq.~(\ref{eq:Zch1}) over ${\bf p}_{\perp}$ yields,
\begin{equation}
\label{eq:intp}
\int d {\bf p}_{\perp} e^{-i {\bf p}_{\perp} \cdot {\bf R} } {\cal Z}_{\psi} \left< e^{-\beta H_{\rm self}} \right>_{\psi} \approx (2 \pi)^{N_0} \frac{4\pi}{N_s b^2} e^{-R^2/N_s b^2}
e^{ \left<  H_{\rm self} \right>_{\psi} ({\bf p}_{\perp}^*) } ,
\end{equation}
where we use ${\cal Z}_{\psi}$ from Eq.~(\ref{eq:Zpsi}) together with the approximation $J_0(p_{\perp}b) = \exp[\log J_0(p_{\perp}b)] \approx \exp [\log (1-p_{\perp}^2b^2/4)] \approx \exp  (-p_{\perp}^2b^2/4)$, which is justified for the range of $p_{\perp}$, that contribute to the integral. We also use the steepest descent method, justified for $N\gg 1$ and $N_s \gg 1$, to evaluate the integral with the saddle point at ${\bf p}_{\perp}^*= -2 i {\bf R} /N_s b^2$.

Finally, we integrate over $\eta_1, \ldots \eta_N$ in Eq.~(\ref{eq:Zch1}), applying the mean-field approximation,
$$
\sum_{s=s_1}^{s_2} \eta_s \approx \sum_{s=s_1}^{s_2} \left<\eta_s \right> =
|s_2-s_1| \left<\eta_s\right> =|s_2-s_1| \frac {z_{\rm top}}{N},
$$
and using again the steepest descend method in the integration over $p_z$; this eventually leads to the following expression for the conditional free energy:
\begin{equation}
\label{eq:Ffin}
\beta F_{\rm ch}  (N, z_{\rm top}, {\bf R}) = \beta U_{\rm sp}(z_{\rm top}) + \frac{\tilde{l}_B N^2}{\tilde{z}_{\rm top} } (\log N-1)  + p^* \tilde{z}_{\rm top} - W(p^*)\nonumber
\end{equation}
\begin{equation}
+ \frac{R^2}{N_s b^2}  +  \frac{\pi \sqrt{2} b \tilde{l}_B N_s^2 }{R} + W_1(N, z_{\rm top},R)  -\log \pi N_s -N_0  \log 2\pi \nonumber
\end{equation}
where $\tilde{z}_{\rm top}= z_{\rm top}/b$ and  $p^*\simeq \tilde{E} \tilde{z}_{\rm top}$ is the saddle point. We also  define
\begin{eqnarray}
W(p^*) =  \frac{1}{\tilde{E}} \left[ {\rm Ei}(\zeta_0) - {\rm Ei}(\zeta_N) + \log\left| {\zeta_0}/{\zeta_N} \right| \right],
\end{eqnarray}
with ${\rm Ei}(x)$ being the exponential integral function,  $\zeta_0=p^*-\tilde{E}$,  and $\zeta_N=p^*-\tilde{E}N$ and

\begin{equation}
\label{eq:W1}
W_1(N, z_{\rm top},R)=\frac{l_B N N_s}{R}\left[ \log \left(1+\sqrt{1+z^{*\,2}_{\rm top}} \right) + \frac{1}{z^*_{\rm top}} \log \left(z^*_{\rm top} +\sqrt{1+z^{*\,2}_{\rm top}}\right)
-\log z^*_{\rm top}\right]  \nonumber
\end{equation}
\begin{equation}
-\frac{l_B N}{R z^*_{\rm top}} \log (2 N_s z^*_{\rm top}),
\end{equation}
where $z^*_{\rm top} = z_{\rm top}/R$.

The impact of counterions on the conformation of the bulk part of the chain may be estimated as a weak perturbation. Referring for the computational detail to  Ref.~\cite{Bril_nanov3} we give here the final result:
\begin{equation}\label{eq:Fc_ch1}
\beta F_{\rm count} (z_{\rm top}) \simeq -\frac{z_{\rm top}}{2 \mu_{GC}} N,
\end{equation}
where $\mu_{GC} =1/(2 \pi \sigma_c l_B q)$ is the Gouy-Chapman length based on the apparent surface charge density $\sigma_c = qN_0/S$ associated with the counterions and $S$ is the lateral area of the system.

\subsection{Dependence of the force and deformation on the external field}
Now we can determine the dependence on the electric field of the PE dimensions as well as the deformation of target body. Simultaneously one obtains the dependence on applied field of the force that arises between chain and the target body. This may be done minimizing the total free energy of the system $F(N,z_{\rm top}, R) = F_{\rm ch}  (N, z_{\rm top}, R)+F_{\rm count} (z_{\rm top}) $ with respect to $N$, $z_{\rm top}$ and  $R$ and using $N_s=N_0-N$ and the constraint $z_{\rm top} \leq bN$.  This allows to find  $N$, $z_{\rm top}$ and  $R$ as functions of the applied electric field, that is, to obtain $N=N(E)$, $z_{\rm top}=z_{\rm top}(E)$ and  $R=R(E)$. Then one can compute the force acting onto the target body. It reads,
\begin{equation}\label{eq:minztop}
   \tilde{f}(\tilde{z}_{\rm top})= \tilde{E}\tilde{z}_{\rm top}
    -\frac{\tilde{l}_B N^2 (\log N -1)}{\tilde{z}_{\rm top}^2} -\frac{Nb}{2 \mu_{GC}}
+\frac{\partial W_1 }{\partial \tilde{z}_{\rm top}} ,
\end{equation}
where $\tilde{f}=\beta b f(z_{\rm top})$, with $f(z_{\rm top})= - \partial U_{\rm sp} /\partial z_{\rm top}$ being  the reduced force for a particular force-deformation relation. In the above equation we exploit  $p^* = \tilde{E}\tilde{z}_{\rm top}$ and the saddle point equation, $iz_{\rm top} -\partial W (p_z)/\partial p_z=0$,  valid for $p_z=p^*$.

\section{Clenching efficiency of nano-vices }

To quantify the efficiency of clenching by nano-vices we analyse self-diffusion of a particle squeezed by the planes of the device. One can say that the body is effectively kept  at rest if the self-diffusion coefficient of the clenched particle drops down by a few orders of magnitude, as compared to this  value of a free particle.

For simplicity we assume that both planes are kept parallel and that two equal, normal to the plane, forces act on the top and bottom of the particle. We neglect twisting and rolling motion of the clenched particle and consider only sliding, that is, the tangential  motion. When the particle moves tangentially, it experiences the following set of forces: (i) A viscous force from the surrounding fluid, ${\bf F}_{\rm vis} = -\gamma {\bf v}$, that acts against the velocity ${\bf v}$ with the friction coefficient $\gamma= 6 \pi \eta R_p$, where $\eta$ is the fluid viscosity and $R_p$ is the radius of the particle. (ii) A random force from the surrounding fluid $\vec{ \xi}(t)$, related to the viscous friction. We assume that this is a $\delta$- correlated force, that obeys the fluctuation-dissipation theorem:
$$
\left< \xi_i(t) \xi_j(t^{\prime}) \right> = 2k_BT \gamma \delta_{ij} \delta(t-t^{\prime})
\qquad \quad i,j = x,\, y\, z.
$$
(iii) A "solid friction" force between the planes and the particle ${\bf F}_{\rm sol.fr}$.
Hence one can write the stochastic equation of motion for the Brownian particle between the planes:
\begin{equation}
\label{eq:Lang}
m\frac{d^2 {\bf r}}{dt^2} +\gamma \frac{d {\bf r}}{dt} = {\bf F}_{\rm sol.fr} + \vec{\xi}(t).
\end{equation}

\begin{figure}
\center{\includegraphics[width=0.7\linewidth]{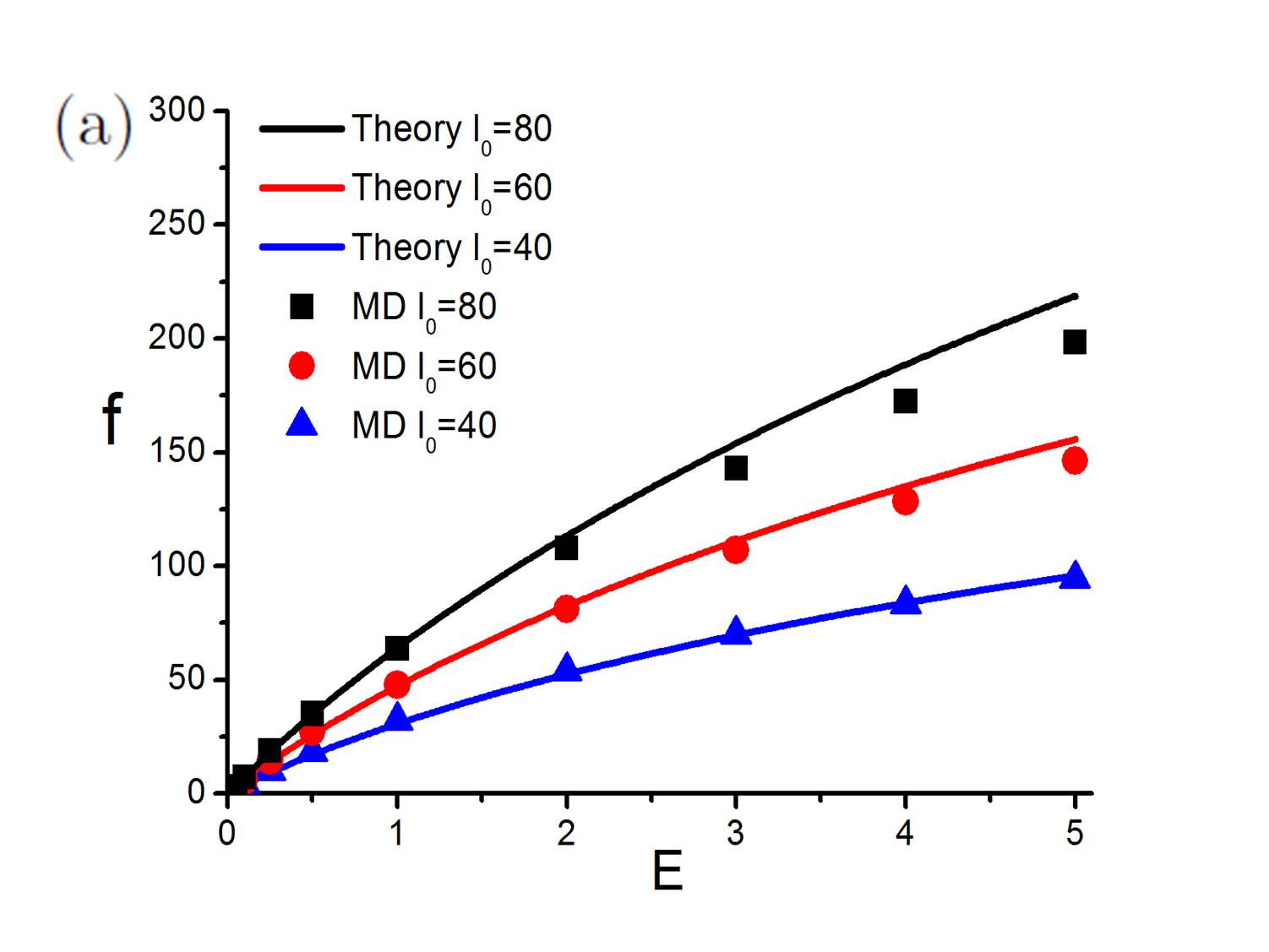}}
\center{\includegraphics[width=0.7\linewidth]{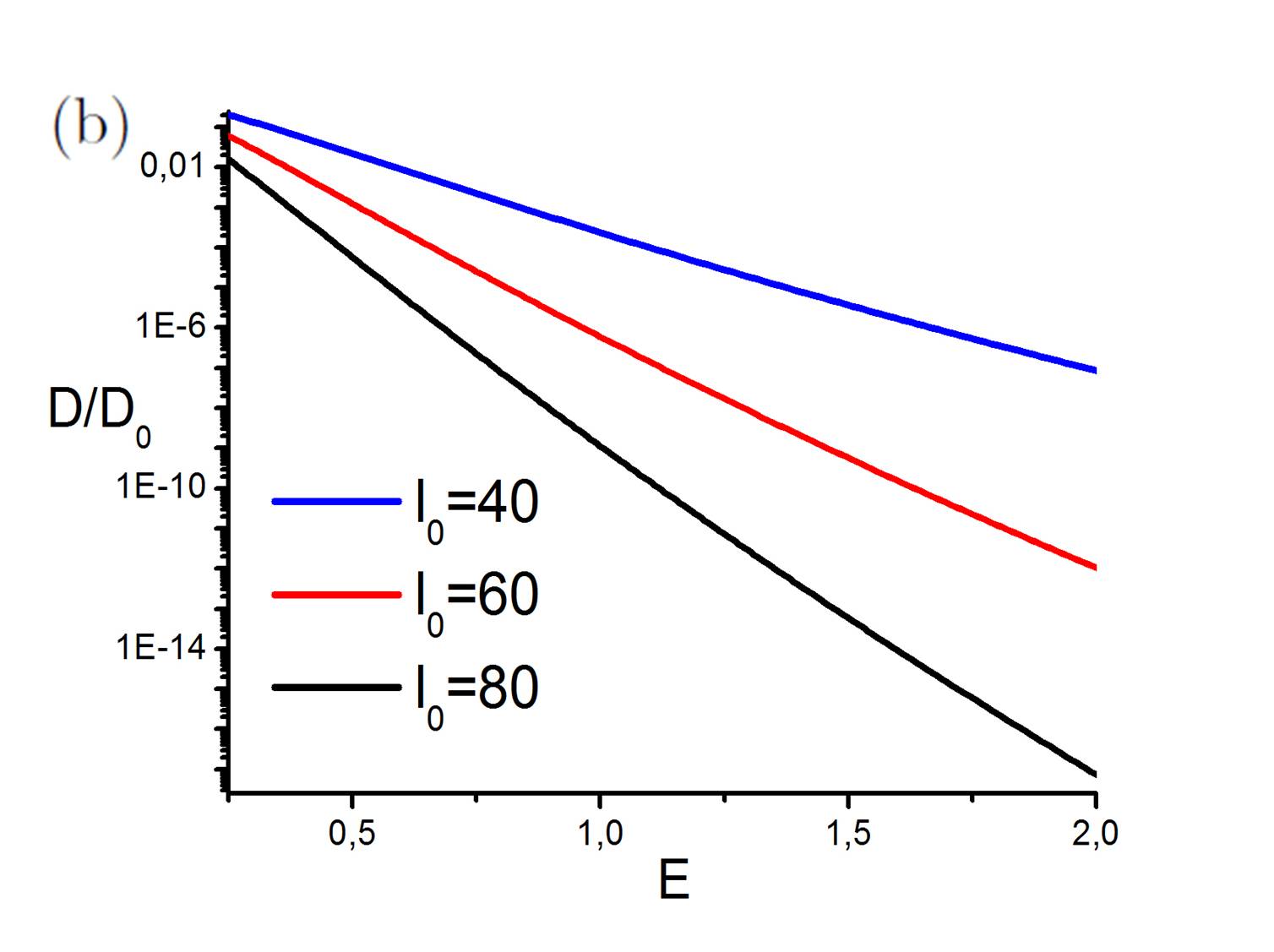}}
\caption{(a) The reduced force generated by the external field, $\tilde{f}=fb/k_BT$, as a function of the reduced electric field, $\tilde{E}=qeEb/k_BT$. The equilibrium length of the Hertzian spring, that corresponds to the diameter of an undeformed nano-particle is $l_0=40b$, $60b$ and $80b$. The reduced force constant is $\tilde{\kappa}= kb^{5/2}/k_BT=1$. The length of the chain is $N_0=320$. (b) The dependence of the diffusion coefficient of a clenched nano-particle $D/D_0$, on the external reduced field  $\tilde{E}$ for different particles sizes. (For a better visibility $\tilde{f}$ and $\tilde{E}$ are shown on the axes without tildes)}
\label{fig3}
\end{figure}

The microscopic derivation of the force ${\bf F}_{\rm sol.fr}$ is rather challenging, therefore we exploit here the Coulomb friction model in the microscopic interpretation of
Cundall and Struck~\cite{CundallStruck}. Note that this model has been used for the tangential friction between colloidal particles~\cite{BB2008}. In the Cundall-Struck model it is assumed that the solid friction force, that counteracts an  externally applied  force is equal to a harmonic spring  force $-\kappa \Delta$, where $\Delta=({\bf r}- {\bf r}_0)$ and ${\bf r}_0$ is the initial position of a body, before the external force started to act. After the displacement from the initial position reaches some quantity $d$, the springs breaks down and the body remains displaced by $\Delta =d$ in the direction of the acting force. For the next action of the external force the new initial position ${\bf r}_0$ corresponds to the shifted one. The maximal force in this model is $\kappa d$, while the average force is equal to $(1/d) \int_0^d \kappa x dx = \kappa d/2$. On the macroscopic scale $L \gg d$ the body moves smoothly with the average resistance force equal to $F_{\rm sol.fr} = \mu F_{\perp}=\kappa d/2$, which yields, $\kappa =2\mu F_{\perp}/d$. One can also write the solid friction force, in the regime of acting spring as the derivative of the according potential,  $F_{\rm sol.fr} = - \nabla U $ with $U= \mu F_{\perp} ({\bf r}- {\bf r}_0)^2/d$. Since the solid friction force always acts against the applied force, we can write the above Eq.~(\ref{eq:Lang}) as the one-dimensional equation in the direction of the applied stochastic force ${\bf \xi}$. Moreover, if we assume that the viscous force is large, $m/\gamma \ll 1$, we recast the above Langevin equation into the overdamped form:
\begin{equation}
\label{eq:Lang_over}
\dot{x} =\gamma^{-1} \frac{d U}{dx} + \tilde{\xi}(t)
\end{equation}
with $\left< \tilde{\xi}(t) \tilde{\xi}(t^{\prime}) \right> = 2D_0 \delta(t-t^{\prime})$, where $D_0=k_BT/\gamma$ is the diffusion coefficient  of a non-constrained particle.

Note, that in contrast to conventional overdamped Langevin equation, the above Eq.~(\ref{eq:Lang_over}) describes the potential $U(x)$, centered after each jerk of size $d$, at a new position. Hence the random tangential  motion of a particle occurs as following: If a displacement of the particle due to the action of the stochastic force $\xi(t)$ does not exceed $d$, the particle performs a Brownian motion in the harmonic potential $U(x)$, centered at ${\bf r}_0$,  with zero self-diffusion coefficient. If at some moment $t_0$ the displacement becomes equal to ${\bf d}$, the particle shifts to ${\bf r} = {\bf r}_0 +{\bf d}$ and starts to perform Brownian motion in the potential $U(x)$, centered at ${\bf r} = {\bf r}_0 +{\bf d}$. Let the critical displacements by vectors ${\bf d}_0$, ${\bf d}_1, \ldots$, ${\bf d}_k$, occur at times $t_0$, $t_1, \ldots$, $t_k$. Due to the Markovian properties of the random force $\xi(t)$ all the vectors ${\bf d}_k$ and the instances $t_k$ are independent. Moreover, we can assume that the sequence of times $t_1, \ldots$, $t_k$ obeys the Poisson distribution
$$
P_k(t) = \frac{1}{k!} \left(\frac{t}{\tau} \right)^ke^{-t/\tau} \, ,
$$
where $\tau$ is the average time between the successive jerks. Then the mean square displacement during time $t$ reads,
\begin{eqnarray}
\label{eq:MSD}
\left< \Delta {\bf r}^2(t) \right> &=&\sum_{k=0}^{\infty} k d^2 P_k(t) =
d^2 \sum_{k=0}^{\infty} k  \frac{1}{k!} \left(\frac{t}{\tau} \right)^ke^{-t/\tau}  =  \left(d^2/\tau \right) t = 2 Dt\, ,
\end{eqnarray}
which implies that the diffusion coefficient is $D=d^2/2 \tau$. In the above equation we take into account that $\left< {\bf d}_i \cdot {\bf d}_j \right> = \delta_{ij} d^2$, due to independence of the displacements ${\bf d}_i$ and $ {\bf d}_j$ for $i \neq j$.

Hence, to find the diffusion coefficient one needs to compute the average waiting time $\tau$. This may be done if we find the so-called "mean-first-passage time" for the potential $U(x)$ \cite{Gardiner,Redner}.  The equation for the average time $T(x)$, needed for a particle, initially located at a point $x$, within a potential well $U(x)$, to reach the point $x=\pm d$, where the jerk takes place, reads \cite{Gardiner}:
\begin{eqnarray}
\label{eq:MFPT}
&&D_0\frac{d^2 T(x)}{d x^2 } - \gamma^{-1} \frac{d U(x)}{d x }\frac{d T(x)}{d x }+1=0 \nonumber \\
&& T(-d) =T(d)=0
\end{eqnarray}
where $U(x) = \mu F_{\perp} x^2/d=\alpha (x^2/d^2)$, with  $\alpha = \frac{\mu F_{\perp}d}{k_BT }$ and we choose ${\bf r}_0=0$. The boundary conditions at $x=\pm d$ are obvious. The solution to Eq.~(\ref{eq:MFPT}) may be expressed in terms of the hypergeometric function $H(z)={}_2F_{2} (1,1; 3/2,2; z)$:
\begin{equation}
\label{eq:Tx}
T(x)=\frac{d^2}{2D_0} \left[ H\left( \alpha \right) - \frac{x^2}{d^2} H\left(\alpha \frac{ x^2}{d^2} \right) \right].
\end{equation}
Now we average $T(x)$ over the starting point $x$, using the equilibrium distribution of starting points within the potential well,  $ \sim e^{-\beta U(x)}$. The results then read,
\begin{eqnarray}
\label{eq:tau_av}
\tau &=& \left< T(x) \right>_{\rm eq}
=\int_{-d}^d  Ce^{-\beta U(x)} T(x) =\frac{d^2}{2D_0} \left[ H\left( \alpha \right) + \frac{e^{-\alpha}}{\sqrt{\pi \alpha} {\rm Erf}(\sqrt{\alpha})} -\frac{1}{2 \alpha} \right], \nonumber
\end{eqnarray}
where $C^{-1}=\int_{-d}^d e^{-\beta U(x)} dx$ is the normalization constant and in the last expression we expand $H(x)$ around $x=0$ and keep only the leading term.
If the potential well is deep enough, that is, $\beta U(d) \gg 1$, then the above expression may approximated by its asymptotic value for $\alpha \gg 1$. Taking into account that $H(x^2) \simeq \sqrt{\pi} e^{x^2}/2 x^3$ for $x \gg 1$ and using Eq.~(\ref{eq:MSD}) we find the effective self-diffusion coefficient:
\begin{eqnarray}
\label{eq:tauappr}
D(E)=\frac{2 D_0}{\sqrt{\pi}} \left(\frac{\mu F_{\perp} d}{k_BT} \right)^{3/2} e^{-\mu F_{\perp} d/k_BT}.
\end{eqnarray}
Here $D$ depends on the field $E$ since $F_{\perp}=F_{\perp}(E)=2f(z_{\rm top})$, that  has been computed in the previous section. Using the value of $\mu = 0.2$, which is motivated by the friction coefficients between polymer surfaces (e.g. $\mu=0.5$ for polystyrene-polystyrene, $\mu =0.15-0.25$ for
nylon-nylon, etc.), and taking into account that the friction force acts on the top and bottom part of a clenched particle, which duplicates the friction force, we calculate $D$ as a function of the electric field.  The results for $D(E)$ are shown in Fig.3(b). For simplicity in our calculation we use $d=b$ for the microscopic parameter of the Cundall-Struck friction model.
\section{MD simulations}
%
The MD simulations have been performed for a freely joint
bead-spring chain. All $N_0=320$ beads, except to the one, bound to the plane at $z=0$,  carry one (negative) elementary charge. $N_{0}$ monovalent free counterions of opposite charge make the system electroneutral.  We assume the implicit good solvent which implies short-ranged, purely repulsive interaction between all particles, described by a shifted Lennard-Jones potential. Neighboring beads along the chain are connected by the standard
FENE potential, e.g.~\cite{CSA00,KUM05}. The  bond length at zero force is  $b \simeq \sigma_{LJ}$ with $\sigma_{LJ}$ being the Lennard-Jones parameter. All particles except the anchor bead interact with a short-ranged repulsive potential with the grafting plane at $z=0$ and upper plane at $z=L_{z}$. The charged particles interact with each other and the external field with unscreened Coulomb potential, quantified by the Bjerrum length. We set  $l_B=\sigma_{LJ}$ and use a Langevin thermostat to hold the temperature $k_{B}T/\epsilon_{LJ}=1$, where $\epsilon_{LJ}$ is the Lennard-Jones energy parameter. More simulation details are given in Refs.~\cite{CSA00,KUM05}. The free end of the chain is linked to a deformable target body, which is modeled by springs with various force-deformation relations. We use the Hertzian force, which can describe the "core" interactions between nano-particles~\cite{Hisao:2010}.  For simplicity, we assume that the anchor of a spring is fixed and  aligned in the direction of the applied field.  The footprint of the simulation box is $L_{x}\times L_{y}  = 424  \times  424 $ (in units of $\sigma_{\rm LJ}$) and the box height is $L_{z}=L = 160$. A typical simulation snapshot is shown in Fig.1(c). We observed that already for relatively weak fields, starting from the field of  about $Eqeb/k_BT  \approx   0.1$ and higher fields, the adsorbed part of the chain is almost flat, while the bulk part of the PE chain in strongly stretched along the field. Moreover, in  sharp contrast to the field-free case~\cite{Winkler98,Brill98,MickaHolm1999} the countertions are decoupled from the PE and accumulate near the upper plane.

\section{Results and discussion}
In Fig.3(a)  we show  the results of  MD simulations and compare them to the theoretical predictions. In particular, the dependence of the field-induced force, acting on the target body is shown as the function of the external electric field.  As it may be seen from the figure a very good agreement between the theory and MD data is observed in the absence of any fitting parameters. Note however, that the theory has been developed for a highly charged chain with a relatively strong self-interactions  and interactions with the charged plane. Some systematic deviations are observed for large fields for the largest nano-particle. This may be probably attributed to a possible elongation of the bond length for strong forces, which is not accounted in the freely joint chain model with the fixed bond length $b$. Note that for aqueous solutions at the ambient conditions, the characteristic units of the force and field are $k_BT/b \approx k_BT/l_B \approx 6\, {\rm pN}$ and $k_BT/be \approx k_BT/l_Be \approx 35\, {\rm V/\mu m}$, respectively;  the latter value is about an order of magnitude smaller than the critical breakdown field for water. Also note that the range of electric fields roughly corresponds to a surface charge density of $0.1 - 10\,{\rm μC/cm}^2$, typical for charged graphite surfaces~\cite{surf_charge}.

 In Fig.3(b) the dependence of the diffusion coefficient of the clenched particle on the external field is shown. One can see a dramatic decrease of $D$ even for relatively weak fields. The effect becomes even more pronounced for larger particles, where the characteristic field of $\tilde{E} =1$ reduces the diffusion coefficient by  nine orders of magnitude. This proves the effectiveness of the PE-based nano-vices.

\section{Conclusion}
We investigate the generation of a mechanical force by a polyelectrolyte (PE) chain grafted to a plane and exposed to an external electric field, when it is linked to a deformable target body. The MD simulations are performed and an analytical theory of this phenomenon is elaborated.  The focus is made on the case, when the force-deformation relation corresponds to that of a Hertzian spring, which quantifies repulsive interactions between nano-size objects, like colloidal particles. The theoretical dependencies for the generated force, acting on the target body, on the external field are in a very good agreement with the simulation data.

Based on the developed theory of the force-to-field response we analyse the efficacy of PE-based nano-vices, comprised of two planes connected by PEs and exposed to an external electric field. By varying the electric field one can clench and unclench a nano-size particle  placed between the planes. We apply Cundall-Struck solid friction model to describe the friction between a particle and clenching planes and develop a theory for self-diffusion coefficient of a clenched particle.  It is shown that the self-diffusion coefficient of a clenched particle drastically decreases even at relatively small electric fields. This proves that PE-based nano-vices may be effectively used to clench nano-size objects as a possible alternative to the existing nano-tube nano-tweezers and optical tweezers. Such devices may find a wide application in future nano-industry, when it is needed to keep a nano-size object immovable, say in various assembly processes. Among important advantages of the discussed nano-vices are (i)~their possibility to operate in aqueous solutions, including solutions with salt, (ii)~a wide range of operating forces and distances, which may be realized within a single device and (iii)~a large variety of molecular structures of PEs that may be used to produce nano-vices.





\begin{thebibliography}{00}

\bibitem{Nanotweez99}
P. Kim and C. Lieber, Nanotube Nanotweezers, Sceince 286  (1999) 2148.

\bibitem{MacDonald2002}
P. MacDonald, et al., Creation and manipulation of three-dimensional optically trapped structures, Science 296 (2002) 1101.

\bibitem{Nanosheets2011}
J. N. Coleman, et al., Two-Dimensional Nanosheets Produced by Liquid Exfoliation of Layered Materials, Science 331 (2011) 568.

\bibitem{Bril_nanov1}
N. V. Brilliantov and C. Seidel, Grafted polyelectrolyte in strong electric field under load: Field-regulated force and chain contraction, Europhysics Lett.,  97 (2012) 28006.

\bibitem{Bril_nanov2} C. Seidel, Yu. A. Budkov and N. V. Brilliantov, Field-regulated force by grafted polyelectrolytes, J. Nanoeng. Nanosystems, 227 (2013) 142.
\bibitem{Bril_nanov3}
N. V. Brilliantov,  Yu. A. Budkov and C. Seidel, Generation of mechanical force by grafted polyelectrolytes in an electric field, Phys. Rev. E 93 (2016) 032505.

\bibitem{Boru98} I. Borukhov   D. Andelman and H. Orland, Scaling laws of polyelectrolyte adsorption, Macromolecules  31 (1998) 1665.

\bibitem{Joanny98} X. Chatellier, and J.-F. Joanny, Pull-off of a polyelectrolyte chain from an oppositely charged surface, Phys. Rev. E 57 (1998) 6923.

\bibitem{Muthu2004} M. Muthukumar, Theory of counter-ion condensation on flexible polyelectrolytes: adsorption mechanism, J. Chem. Phys. 120 (2004) 9343.

\bibitem{Dobry2000} A. V. Dobrynin,   A. Deshkovski, and M. Rubinstein, Adsorption of polyelectrolytes at an oppositely charged surface, Phys. Rev. Lett.  84 (2000) 3101.

\bibitem{Borisov2001} O. V. Borisov, et al., Polyelectrolytes tethered to a similarly charged surface, J. Chem. Phys. 114 (2001) 7700.

\bibitem{Netz2003} R.R. Netz, Nonequilibrium Unfolding of Polyelectrolyte Condensates in Electric Fields, Phys. Rev. Lett. 90 (2003) 128104.


\bibitem{MetzlerCherst:2014}
R. G. Winkler and A. G. Cherstvy,
Strong and Weak Polyelectrolyte Adsorption onto Oppositely Charged Curved Surfaces,
Adv. Polym. Sci.,  255 (2014) 1.

\bibitem{MetzlerCherst2015}
S. J. de Carvalho, R. Metzlerb  and A. G. Cherstvy,
Inverted critical adsorption of polyelectrolytes in confinement,
Soft Matter, 11 (2015) 4430.

\bibitem{Budkov_salt}
Yu. A. Budkov, C.Seidel and N. Brilliantov, Grafted polyelectrolytes in electric field under load. Impact of salt concdentration on the response force, (2016) in preparation.


\bibitem{CSA00} F. S. Csajka, and C. Seidel, Strongly Charged Polyelectrolyte Brushes: A Molecular Dynamics Study, Macromolecules 33 (2000) 2728.

\bibitem{KUM05} N. A. Kumar, and C. Seidel, Polyelectrolyte Brushes with Added Salt, Macromolecules 38 (2005) 9341.


\bibitem{Hisao:2010} K. Saitoh,  et al., Negative Normal Restitution Coefficient Found in Simulation of Nanocluster Collisions, Phys. Rev. Lett. 105 (2010) 238001.

\bibitem{CundallStruck}
  P. A. Cundall and O. D. L. Strack, A discrete numerical model for granular assemblies,  Geotechnique 29 (1979) 47. 
\bibitem{BB2008}
V. Becker and H. Briesen, Tangential-force model for interactions between bonded colloidal particles, Phys. Rev. E 78 (2008) 061404.

\bibitem{Gardiner}
C.W.Gardiner, \emph{Handbook of Stochastic Methods,} (Springer, Berlin, 1985), 2nd ed.

\bibitem{Redner}
S. Redner {\sl} \emph{A Guide to First-Passage Processes}, Cambridge University Press (2001).

\bibitem{Winkler98} R. G. Winkler, and M. Gold, and P. Reineker, Collapse of Polyelectrolyte Macromolecules by Counterion Condensation and Ion Pair Formation: A Molecular Dynamics Simulation Study, Phys. Rev. Lett. 80 (1998) 3731.
\bibitem{Brill98} N. V. Brilliantov,  and D. V. Kuznetsov and R. Klein, Chain Collapse and Counterion Condensation in Dilute Polyelectrolyte Solutions,
         Phys. Rev. Lett. 81 (1998) 1433.
\bibitem{MickaHolm1999} U. Micka, and  C. Holm, and K. Kremer, Strongly Charged, Flexible Polyelectrolytes in Poor Solvents: Molecular Dynamics Simulations, Langmuir 15 (1999) 4033.
\bibitem{surf_charge}
M. Fedorov and A. A. Kornyshev, Ionic Liquids at Electrified Interfaces, Chem. Rev. 114 (2014) 2978; S. A. Kislenko et al., Molecular dynamics simulation of the electrochemical interface between a graphite surface and the ionic liquid [BMIM][PF6], Phys. Chem. Chem. Phys. 11 (2009) 5584.




\end{thebibliography}
\end{document}